\documentclass{article}[12pt]
\usepackage{amsmath,amsthm,amssymb}

\newcommand{\beq}{\begin{equation}}
\newcommand{\eeq}{\end{equation}}
\newcommand{\bea}{\begin{eqnarray}}
\newcommand{\eea} {\end{eqnarray}}

\def\un\a{{\underline\alpha}}

\def\a{{\alpha}}

\def\Tr{{\mathrm{Tr}}}

\begin{document}

\title{Distinguishing initial state-vectors\\ from each other in histories formulations\\ and the PBR argument}

\author{Petros Wallden\footnote{1. SUPA, Physics Department, Heriot-Watt University, Edinburgh EH14 4AS, United Kingdom 2. Nuclear \& Particle Physics Section, Physics Department, University of Athens, Panepistimiopolis 157-71, Athens, Greece. Email: petros.wallden@gmail.com; petros.wallden@hw.ac.uk}}

\maketitle
\begin{abstract}
Following the argument of Pusey, Barrett and Rudolph \cite{PBR}, new interest has been raised on whether one can interpret state-vectors (pure states) in a statistical way ($\psi$-epistemic theories), or if each one of them corresponds to a different ontological entity. Each interpretation of quantum theory assumes different ontology and one could ask if the PBR argument carries over. Here we examine this question for histories formulations in general with particular attention to the co-event formulation. State-vectors appear as the initial state that enters into the quantum measure. While the PBR argument goes through up to a point, the failure to meet some of the assumptions they made does not allow one to reach their conclusion. However, the author believes that the ``statistical interpretation'' is still impossible for co-events even if this is not proven by the PBR argument.
\end{abstract}

\section{Introduction}

In quantum theory the state of a system is represented by the wavefunction \footnote{A deeper discussion, could involve questioning the whole concept of wavefunction and Hilbert space as starting point of quantum theory, but this is beyond the scope of this paper.}. While there is general agreement on how to use this state in order to extract predictions in the form of relative frequencies of outcomes of multiple copies, there is strong debate on the meaning of the wavefunction for single systems and of the interpretation that is given to it. On the one hand, one can claim that the state represents some real properties of the (single) system and thus attain an ontological status. On the other hand, one can claim that the state reflects the experimenters information about properties of the system. In the latter view, the state is understood as a statistical distribution of different potential realities (if one assumes that it makes sense talking about the real properties of the system). Here we should stress, that the above considerations, concern \emph{pure} states, or in other words state-vectors, and in the following text when we use the term ``state'' with no further specification, it should be understood as a state-vector.

\subsection{Statistical distribution or distinct ontological entities?}

If one was to adopt the statistical view of the state, one would be lead to conclude the following: There is no way with a single experiment to be able to deduce with 100\% certainty which of two non-orthogonal states was your initial state. This is in analogy with classical statistical physics, in the following sense. If the experimenter has one of two possible distributions as initial information (corresponding to two different initial states), and the distributions are overlapping, there is no single experiment that can determine with certainty, which of the two initial distributions was correct. This is due to the fact, that reality corresponds to a single point, and since the distributions are overlapping, there are some potential realities that is consistent with either initial distributions/states.

If on the other hand, one takes the view that the state reflects some ontological property of the system, then one should be able in principle to construct a series of measurements (for the single system considered) that would be able to distinguish between any two non-orthogonal states. This statement concerns the wavefunction of single system and not about some statistical distribution of the relative frequencies of outcomes of many identically prepared copies. Here we should note, that in principle one can maintain the opinion that different states corresponds to different realities, even if we cannot possibly distinguish them experimentally. However, what is certain is that if one \emph{can} distinguish between any two states, then it is difficult to maintain the view that the state corresponds to a statistical distribution.

The argument of Pusey, Barrett and Rudolph (from now on referred to as PBR) \cite{PBR}, provides an algorithm distinguishing between non-orthogonal states and thus (according to the authors) ruling out a statistical interpretation or what they call ``\emph{$\psi$-epistemic theories}''. In \cite{PBR Experiment} the argument has been tested experimentally, and following the assumptions that they made and re-stated, quantum theory is confirmed and $\psi$-epistemic theories ruled out. The argument along with its assumptions, will be presented in detail later.

The specifics of the PBR argument was claimed to be independent of the actual ontology of quantum theory and thus potentially independent of the interpretation one chooses. However it would be interesting to examine the arguments details for particular interpretations and how the assumptions made carry over. Particular interest, would be to do so for alternative formulations of quantum theory (where the argument does not follow trivially) such as the histories formulation, and the relation this has with the ability to retrodict properties of the initial state of the universe. This precisely is the topic of this paper.

\subsection{Histories and initial state}

In order to do this comparison, one first needs to understand what is the meaning and role both conceptually and mathematically, of state-vectors in histories formulations. Here, we should specify what we mean as histories formulation. They are formulations of quantum theory, that assign (or use) a quantum amplitude or a quantum measure to histories of the system. In other words, formulations based on the Feynman path integral\footnote{One can view the De Broglie-Bohm theory as a histories formulation, however in this paper we use the term for formulations based on the Feynman path integral and the decoherence functional, which is not the case for De Broglie-Bohm}. Examples of full interpretations based on the path integral, are the decoherent (or consistent) histories approach (e.g. \cite{Decoherent Histories}) and the co-event formulation (e.g. \cite{co-events,coevents_review}) which we will introduce in section 3.

In these formulations, the concept of the state of the system at a (random) moment of time does not make sense. The only place that the state enters the picture is the \emph{initial} state of the universe strictly speaking or more practically, the initial state of a subsystem we consider\footnote{This is achieved by restricting the path integral to particular sets of histories and weighting differently each of those}. In the latter case, the initial state, represents the complete summary of the past of our system. Mathematically it enters by modifying the amplitudes of histories and thus the decoherence functional and the quantum measure (see below).

In histories formulations, depending on the particular interpretation, one gives ontological status to either a single history or a subset of histories (or to something else such as a co-event which will be defined properly in section 3.3) but \emph{not} to the state of the system itself. However, one is still able to ascribe ontological status to the initial state in the following sense. Since we are not doing deterministic physics, starting from some initial state gives rise normally to several potential realities. If one can show that the set of potential realities that arise, if we start with one initial state is completely disjoint with the set of potential realities of any other distinct initial state, then by determining the actual realised reality, we can retrodict uniquely the state we started from. One can speak about properties of the initial state which, in this sense, attains ontological status. In other words, the universe where the initial state is $|\Psi_1\rangle$ is a different one from one that has a distinct initial state $|\Psi_2\rangle$.

If on the other hand, one wishes to interpret the initial state as some short of statistical distribution, then it is necessary that for a given reality there are more than one initial states that are compatible with. In this case, it would be impossible, no matter how fine-grained description one has, to have completely disjoint sets of potential realities corresponding to (non-orthogonal) distinct state-vectors.

The procedure to disprove the latter view, is to consider some sufficient fine description, i.e. a sequence of measurements\footnote{Note that measurements are not understood as actual measurements with an external experimenter, but rather by introducing different projections at different times and mathematically fine graining histories. It is always possible however, to view the histories space differently, in terms of the finest possible description and its coarse-grainings, that does not require the concept of projection operator as a prerequisite.} at suitable moments of time that would give rise to sets of potential realties that are disjoint for distinct state-vectors. This will be done in section 4 and in the Appendix.

Here, we should stress, that to draw conclusions for the histories formulations, one needs to have a specific interpretation (what are those potential realities) and is not possible to do so fully, only by looking at the quantum measure.

If one \emph{is} able to distinguish between different state-vectors, that would have an important consequence for histories. We would be able, at least in principle, to retrodict the initial state (if pure), uniquely. This is of great importance, e.g. for cosmological considerations, where the histories formulations have been applied extensively.

\subsection{This paper}

In this paper, in section 2 we will review the PBR argument illustrated with the simplest example of the $|\Psi_1\rangle=|0\rangle,|\Psi_2\rangle=|+\rangle$ states, and point out the assumptions made. In section 3 we will present the histories formulation, by introducing the decoherence functional and the quantum measure and briefly mentioning the co-event and the decoherent histories formulations. In section 4 we analyse the PBR argument for histories, looking the quantum measure and examining the potential co-events in detail for two versions of the example considered in section 2. In section 5 we will see how the assumptions of PBR argument affect our conclusion for the co-events formulation and in section 6 we will summarise and conclude.

\section{The PBR argument}\label{section PBR}

In this part, following \cite{PBR}, we provide an algorithm that one could follow to show that two non-orthogonal state-vectors correspond to distinct realities and could not possibly be confused as a statistical interpretation would imply.

Assume that two states\footnote{Where $\langle\Psi_1|\Psi_2\rangle\neq0$}  $\Psi_1,\Psi_2$ correspond to a statistical distribution of some underlying true properties. This would imply that with some probability $p\neq0$ the true properties of the system are such that they are compatible with both the system being in state $\Psi_1$ and $\Psi_2$. Now imagine we consider a pair of identical systems such that each of them can be either in state $\Psi_1$ or in state $\Psi_2$. This can be realised by considering two boxes that each of them generates either $\Psi_1$ or $\Psi_2$ with some probability $p$. Then, provided that the two systems are independent, with probability $p^2$ the true underlying properties of the composite system would be compatible with the system being in any of the four states\footnote{Note that the above can be generalised for $n$ copies of the system.} $|\Psi_1,\Psi_1\rangle, |\Psi_1,\Psi_2\rangle, |\Psi_2,\Psi_1\rangle, |\Psi_2,\Psi_2\rangle$. Then by choosing to make a measurement in a suitable basis for the composite system, one can show that no outcome of that measurement is compatible with all of those potential initial states.

Let us see it here with a simple example, of two particular non-orthogonal states of a qubit, the $|\Psi_1\rangle=|0\rangle$ and the $|\Psi_2\rangle=|+\rangle=1/\sqrt2(|0\rangle+|1\rangle)$ states.

If $|\Psi_1\rangle\textrm{ and }|\Psi_2\rangle$ were not distinct entities but corresponded to statistical distribution of some underlying true properties, then there must be some possible
realities, where one cannot distinguish between the four states $|00\rangle,|0+\rangle,$ $|+0\rangle,|++\rangle$. In particular the chance of being in one such case is $p^2$ where $p=|\langle0|+\rangle|^2=1/2$. For those cases, no measurement should be able to distinguish between the four above states. In other words, it should not be possible to measure the composite system and get with certainty (probability 1 or 0) that any one of the four above states is not possible, since that would not be compatible with the cases that appear with probability $p^2\neq0$ that those four states are indistinguishable.

Assume now that we measure the composite system in the following basis\footnote{$|-\rangle=1/\sqrt2(|0\rangle-|1\rangle$)}:

\bea\label{basis} |\xi_1\rangle=\frac1{\sqrt2}(|01\rangle+|10\rangle) & , & |\xi_2\rangle=\frac1{\sqrt2}(|0-\rangle+|1+\rangle),\nonumber\\
|\xi_3\rangle=\frac1{\sqrt2}(|+1\rangle+|-0\rangle) & , & |\xi_4\rangle=\frac1{\sqrt2}(|+-\rangle+|-+\rangle)
\eea
We are lead to paradox because, there is no outcome of this measurement that is compatible with the system being \emph{all} four initial states. In particular, if the initial state was $|00\rangle$ then $|\xi_1\rangle$ never occurs, if $|0+\rangle$ then $|\xi_2\rangle$ never occurs, if $|+0\rangle$ then $|\xi_3\rangle$ never occurs and if $|++\rangle$ then $|\xi_4\rangle$ never occurs. But since $\{|\xi_i\rangle\}$ forms a complete basis, one outcome always occurs, and thus with certainty we can conclude that the system never is compatible with all four states $|\Psi_i,\Psi_j\rangle$. Depending on the outcome, every time we rule out one of the four initial states. From this observation, the authors conclude that the single system is not compatible with both $|\Psi_1\rangle$ and $|\Psi_2\rangle$ and thus a statistical interpretation is not possible.

The argument can be extended for any two non-orthogonal states provided one considers a suitable number $n$ of identical (and non-interacting) copies and performs a measurement in a particular basis on the total tensor product Hilbert space. The reader is referred to the original papers \cite{PBR},\cite{PBR Experiment} for details.

 At this point, we should stress what assumptions were made (and stated) by the authors in order to reach their conclusion.
\begin{enumerate}

 \item The quantum states are prepared in isolation of the rest universe and after that the (individual) system has a well defined set of physical properties. Each sub-system is in pure state, so no complications arise from consideration of entangled systems.

 \item It is possible to prepare multiple (identical) systems that are uncorrelated. This can be realised by considering spacelike separated apparatuses or using the same apparatus at much later times. %NOT same at histories...
     Note, that due to this point we are allowed to deduce that if with probability $p$ something occurs in the single system then we get with probability $p^2$ this thing occurring at both copies of the composite system.

\item The measuring devices respond only to physical properties of the measured systems. However, this does not need to be in a deterministic way.

\end{enumerate}
The experimental test of the argument in \cite{PBR Experiment} also assumes the above assumptions, and in particular failure to satisfy them (as we will see in section 5), allows for reaching different conclusion. Further discussion of the assumptions will follow at section 5, and in relation with histories formulations.

\section{Histories and co-events}

In histories formulations the central mathematical structure of interest, is the history space $\Omega$, the space of all finest grained descriptions\footnote{Note, that following Hartle and Sorkin, we are adopting the path integral view of histories that takes the stance that there exist a unique (preferred) fine grained description, i.e. paths in the generalised configuration space. Other points of view, such as Isham's, are compatible with the one we take, at least in most ordinary cases.}. It is the set of all possible histories, and each element of it $h_i\in\Omega$ corresponds to a full description of the system, specifying every detail and property. For example, a fine grained
history gives the exact position of the system along with the specification of any internal degree of freedom, for every moment of time. For a single non-relativistic particle, $\Omega$ would be the space of all trajectories in the physical space. Subsets of $\Omega$ are called events and correspond to all the physical questions one can ask. If, for example, one wishes to ask ``was the system at the region $\Delta$ at time $ t$?'', it corresponds to the subset
$A$ defined as $\{A: h_i\in A \textrm{ iff } h_i(t)\in\Delta\}$, i.e. all histories that the system at time $t$ is in the region $\Delta$.

\subsection{Amplitudes, decoherence functional \\and quantum measure}

One can assign an amplitude (complex number) to each history following the Feynman path integral approach. This amplitude, depends on the initial state and on the dynamics of the system encoded in the action $S$:

\beq\label{Feynman} \a(h_i)=\exp i S(h_i)\eeq
Using this amplitude one can recover the transition amplitudes from $(x_1,t_1)$ to $(x_2,t_2)$ by summing through all the paths $\mathcal P$ obeying the initial and final condition:

\beq \a(x_1,t_1;x_2,t_2)=\int_\mathcal P \exp(iS[x(t)])\mathcal D x(t)\eeq
The mod square of this amplitude is the transition probability. Using the Feynman amplitudes Eq. (\ref{Feynman}) one can define the decoherence functional:

\bea D(A,B)&=&\int_{A} \exp{(-iS[x(t)])}\mathcal{D}x[(t)]\int_{B} \exp{(+iS[y(t)])}\mathcal{D}y[(t)]\times\nonumber\\& &\delta(x(t_f)-y(t_f))\rho(x(t_0),y(t_0))\eea
Where $A$ and $B$ are any subsets of $\Omega$, $t_f$ is the final while $t_0$ the initial moment of time considered and $\rho$ the initial state. The decoherence functional obeys the following conditions:

\begin{enumerate}
\item Hermiticity: $D(A,B)=D^*(B,A)$
\item Bi-linearity: $D(A\sqcup B, C)=D(A,C)+D(B,C)$
\item Strong positivity\footnote{Initially this requirement was weaker, namely that $D(A,A)\geq 0$ and was called positivity. However, it is believed that this stronger requirement is more physical and we will adopt this convention here.}: the matrix $D(A_i,A_j)$ is positive for any collection $\{A_1,A_2,\cdots,A_n\}$ of subsets of $\Omega$.
\item Normalisation\footnote{The sum in the expression should be replaced with an integral if we consider continuous histories}: $\sum_{i,j}D(A_i,A_j)=1$
\end{enumerate}

The decoherence functional can also be defined using time ordered strings of projection operators. In particular

\beq\label{operator df} D(A,B)=\Tr(C^\dagger_A\rho C_B)\eeq
Where $C_A$ and $C_B$ are the class operators, which are strings of time ordered projection operators corresponding to the histories $A$ and $B$ respectively that we will specify below and is defined in the following way.

\beq C_A= P_{A_n}U(t_n-t_{n-1})\cdots U(t_3-t_2)P_{A_2}U(t_2-t_1)P_{A_1}\eeq
Where the $U(t)$ is the unitary evolution operator that relates to the Hamiltonian via $U(t)=\exp (-iHt)$, and $P_{A_i}$ is the subspace that history $A$ lied at time $t_i$. The history $A$ is the subset of $\Omega$ that contains all the histories that the system lies in the subspace that $P_{A_1}$ projects to, at time $t_1$ \emph{and} in the subspace that $P_{A_2}$ projects to, at time $t_2$, etc. The projection operators in general can be at any subspace of the Hilbert space. Note that the expression for the class operator, is precisely the one used in ordinary quantum mechanics to obtain the amplitude, if some external observer carried out those measurements at the given times. By the linearity property of the decoherence functional the Eq. (\ref{operator df}) can be extended to subsets of $\Omega$ that are not just strings of projection operators (called inhomogeneous histories). Here we are not going into deeper discussion of the differences of those two definitions and their interpretational consequences. We will simply use the operator expression for finite moments of time histories, since it is more easy to deal with. However, one can see that the two definitions are essentially equivalent for the examples considered in this paper.

From the positivity condition, we can see that the diagonal elements of the decoherence functional are non-negative. Those terms are also referred to, as quantum measure (\cite{quantum measure}) and are labelled as $\mu(A):=D(A,A)$. It is important to note, that only the \emph{real} part of the decoherence functional affects the quantum measure. This property, in relation with extending the quantum measure for composite systems, will be discussed at section 5.

One could be tempted to interpret the quantum measure as probability. However this is not possible, due to interference. The additivity condition of probabilities, is not satisfied:

\beq \mu(A\sqcup B)\neq\mu(A)+\mu(B)\eeq
However a weaker condition holds that shows that there is no three-paths interference:

\beq \mu(A\sqcup B\sqcup C)=\mu(A\sqcup B)+\mu(A\sqcup C)+\mu(B\sqcup C)-\mu(A)-\mu(B)-\mu(C)\eeq
Interpreting the quantum measure, is the issue of histories formulations, and in this paper we will focus on the decoherent histories and the co-event formulations\footnote{Other attempts have been made, such as using some special precluded sets, at the original papers introducing the quantum measure \cite{quantum measure} by Sorkin, or extending the concept of probability to maintain a single-history-realised view, more recently in \cite{GeHa2011} by Gell-Mann and Hartle}.

Provided the initial state is pure, we can define the quantum measure in terms of amplitudes for histories

\beq \mu(A)=\sum_{i,j}\a^*(h_i)\a(h_j)\times \delta(h_i(t_f)-h_j(t_f))\eeq
It is important to note the delta function that guarantees that histories that do not end (final time $t_f$) at the same point, do not interfere. In other words, the quantum measure becomes additive if one considers alternatives that differ at the final moment of time.

Another thing to note, when one considers the operator definition of the decoherence functional, is what happens if one introduces more moments of time. This is simply a fine graining of the previous histories as it is easily seen from the definition using the paths/trajectories.

A final issue to discuss, before introducing interpretations of the quantum measure, is the sets of histories that have quantum measure zero. Those sets of histories will be referred to as precluded sets. In general, there are two kind of quantum measure zero sets. The trivial ones, that all their subsets have also quantum measure zero (similarly with classical measure zero sets), and the non-trivial, that have subsets that are non-zero. The latter are due to interference and are the source of any counter-intuitive property.

If a set $A$ has quantum measure zero then it decoheres with its compliment\footnote{This is not necessarily true if one considers the weaker notion of positive decoherence functionals.} $\mu(A)+\mu(\neg A)=1=\mu(A\cup\neg A)$. This along with other considerations lead us to the conclusion that quantum measure zero sets, do not occur in nature. We further need to assume that any set $B\subseteq A$ also does not occur if $A$ does not occur, if one wishes to maintain classical deductive reasoning such as the Modus Ponens (see \cite{MP}). However, this could lead us to trouble, because it is known that generally, one can cover the full history space $\Omega$ with zero quantum measure sets \cite{KS}.

\subsection{Decoherent histories approach to quantum theory}

Decoherent histories (also known as consistent histories) is an approach developed, initially, mainly by Griffiths, Omnes and Gell-Mann and Hartle (eg. \cite{Decoherent Histories}). The decoherence functional, first appeared in relation with this approach. The mathematical aim of the approach is to tell when it is possible to assign probabilities to (coarse-grained) histories of a closed quantum system, or in the language we developed above, when is it possible to assign the quantum measure of a set of histories $A$ as the probability of this set $A$ actually occurring.

Different people have given different motivations for the approach, but the general aim is to be able to reason about a closed system with no reference to observer or an a-priori distinction of microscopic and macroscopic degrees of freedom, or distinction of quantum and classical systems. The field of quantum cosmology which is the ultimate closed quantum system, was one of the motivation for this approach, while the way that classicality emerges and its connection with decoherence, was another.

In order to make the quantum measure into a proper classical measure, one needs to restrict attention to some particular collection of subsets of $\Omega$ rather than the full collection of all possible subsets of $\Omega$. The failure to satisfy the additivity condition can be traced at the off-diagonal terms of the decoherence functional as one can see from the very definition\footnote{Strictly speaking, from the real part of the off-diagonal terms. The role the complex part plays will be briefly discussed later.}. Let us take a partition of $\Omega$ which is defined to be a collection of subsets $\mathcal P_1=\{A^1_1,A^1_2,\cdots,A^1_n\}$ where $A^1_i\cap A^1_j=\emptyset$ and $\cup_i A^1_i=\Omega$. The superscript labels the partition considered, while the subscript labels the different cells of one partition. If for any pair of cells of one partition $A^1_i, A^1_j$ it holds that

\beq\label{decoherence condition} D(A^1_i,A^1_j)=0\textrm{ if }i\neq j\eeq
then the partition is called a \emph{consistent set}. For this partition and any further coarse-graining, the standard rules of probability theory hold. The quantum measure, when restricted to those questions, becomes a classical measure. One would be tempted to assign these probabilities to the coarse-grained histories of the partition.

However, one can consider other partitions, say $\mathcal P_2=\{A^2_1,A^2_2,\cdots,A^2_n\}$. It is possible that this partition also forms a consistent set obeying Eq. (\ref{decoherence condition}). Important thing to note, is that there does not exist, one finest-grained consistent set, that all other consistent sets are simply coarse-graining of that. One is not allowed to make propositions involving sets that belong to separate consistent sets, and thus cannot properly assign probabilities to histories once and for all, but it is dependent (contextual) to the consistent set one considers. Counter-intuitive consequences have been analysed (eg. by Dowker and Kent \cite{DoKe}) and at this point we can only say that one cannot assign probabilities to histories in a classical sense. The minimalist view, is that one could use present records corresponding to one consistent set, to deduce things about other present records related with the same consistent set. In this language, being able to deduce things about the initial state of the universe from present records, would imply the ability to make further present time predictions. Being able to distinguish between different initial state-vectors, which is the discussion of the present paper, lie within this scope.

\subsection{The co-events formulation of quantum theory}
The co-event formulation was developed mainly by Sorkin \cite{co-events,Sorkin2010} and collaborators (e.g. \cite{KS,GaWa,Gud,Ga}). A review can be found here \cite{coevents_review}. It is a more recent attempt, to maintain a realistic picture of closed systems quantum theory and being able to speak about properties of the closed system being possessed objectively. In classical physics, there are three structures if one wishes to use the histories language.

First, the histories space $\Omega$ and the collection of all subsets of $\Omega$ which form a boolean algebra $\mathcal U$, second the space of truth values $\mathcal T=\{1,0\}$ also forms a boolean algebra and finally third the valuation maps $\phi$ which assign a truth value $1,0$ to all questions/subsets of $\Omega$. These maps we call them co-events and in classical physics need to respect the boolean structure of $\mathcal U$ and $\mathcal T$ and be a homomorphism

\bea\phi(A\triangle B)&=&\phi(A)+\phi(B)\nonumber\\
\phi(A\cdot B)&=&\phi(A\cap B)=\phi(A)\phi(B)\eea
One can show that there is a one-to-one correspondence between single histories (points at $\Omega$) and homomorphic co-events, in this way

\beq \phi_h(A)=1\textrm{ if } h\in A\textrm{ and }\phi_h(A)=0\textrm{ if } h\in\neg A\eeq
i.e. when the co-event $\phi_h$ is a characteristic map of $h$. We usually assume that reality is one element (say $h$) of $\Omega$, the one that is actually realised. We can see here, due to the above correspondence, that we could have a dual view, and say that the co-event/characteristic map $\phi_h$ is what is truly realised. The potential realities, thus are all the co-events that correspond to some history $h$ that does not have zero (classical) measure.

In quantum theory the above picture cannot be maintained, due to the fact that we have a quantum measure on $\Omega$. One could weaken the requirement that the maps are homomorphisms and allow them to be non-additive. In particular, we could have multiplicative co-events that

\bea\phi(A\cdot B)&=&\phi(A\cap B)=\phi(A)\phi(B)\nonumber\\
\phi(A\triangle B)&\neq&\phi(A)+\phi(B)\eea
Then one can show that all the multiplicative co-events are to one-to-one correspondence not with single histories but to subsets of histories, in other words they are characteristic maps of non-trivial subsets $A$.

\beq \phi_A(B)=1\textrm{ iff } A\subseteq B\eeq
One can see that if $A$ is neither subset of $B$ nor of $\neg B$, then both $B$ and $\neg B$ are false. This precisely is where the strangeness of quantum theory is encoded. However, if the possible $A$ are small enough, one would expect that all classical questions would be too coarse-grained to intersect non-trivially $A$, and no paradox would appear. One should note, that if a particular co-event corresponds to a characteristic function of a set with a single history, it gives rise to completely classical logic (homomorphism) and in this sense we will refer to it as a ``classical co-event'' even if the system it arises may allow other co-events that are not of this type.

Other than the requirement to be multiplicative, the allowed co-events must (a) be preclusive, i.e. respect that $\mu(A)=0\implies\phi(A)=0$ and (b) be minimal (called primitive), i.e. be as small as possible, providing they obey multiplicativity and preclusivity (e.g. \cite{Ga} for greater detail). The first requirement, uses the quantum measure and it is at this point where the dynamics (Hamiltonian) and initial state of the particular system enters the picture. In order to recover the full probabilistic predictions, one apparently needs to use the quantum measure and resort to the Cournot principle. Cournot's principle, is the following: ``In a repeated trial, an event $A$ singled out in advance, of small measure ($\leq\epsilon$), rarely occurs''. The details on how this can be used to recover the probabilistic predictions of quantum theory (e.g. double slit pattern) can be found in \cite{GaWa}, and a shorter version in section 4.2 of reference \cite{coevents_review}. We should stress however, that ontological status of the state-vector that we examine in this paper, is not directly related with the way of recovering the probabilistic predictions. In PBR argument, for example, the particular  values of probabilities, play no role.

To summarise, the potential realities for the co-event formulation, are the set of all primitive, preclusive and multiplicative co-events. From now on, when we mention co-events or potential realities, we will refer precisely to these.

Two important features of the co-event formulation, are the following. First, that the deductive logical inferences are possible, namely the Modus Ponens rule of inference holds for multiplicative co-events (and only for those!) \cite{MP}. Second, that one has a unique finest grained classical partition. In other words, there exist a finest grained description, such that, no matter which co-event is realised, the resulting (coarse-grained) logic is classical (see appendix of \cite{GaWa}). This is different than in decoherent histories, where one has incompatible consistent sets.

As a final point at this section, we should stress that in general there are many potential co-events, given a quantum measure. This is in analogy with classical stochastic physics and we could claim that quantum theory constitutes generalisation of stochastic physics rather than deterministic. We thus have a set of possible realities $\mathcal C$. Measurements made, narrow down this set of potential co-events and we may or may not be able to narrow it down to the single co-event that is actually realised.

If unsure of which was the initial state, we have many possible sets $\mathcal C_i$ of co-events, each set corresponding to one candidate initial state $\psi_i$. If we carry out a measurement and get a result that is not possible in any of the co-events of one particular set (say for example $\mathcal C_3$), we can safely deduce that the initial state was not the one that has as possible co-events this set ($\psi_3$ in this example). In other words we would be able to deduce things about the initial state of the system (or more ambitiously stated, of the universe).

The reader is referred to the original references for a more complete and detailed presentation. Here we only introduced the necessary material and stressed few things that are important for this paper.

\section{PBR for co-events}

In order to see the PBR argument for co-events one has to first construct the above experiment in a histories version, then compute the quantum measure and finally find the potential realities, i.e. the potential co-events. The first two steps are common in all histories formulations, however the conclusion of the argument is possible only after one considers the potential realities.

Different initial states give rise to different quantum measures and thus different set of possible realities/co-events. The PBR argument, needs to show that there exists no common co-event,
between the different sets of possible co-events corresponding to different initial states. If this is shown we will be able to conclude that there is \emph{no possible reality/co-event} compatible with the state being both $
|\Psi_1\rangle$ and $  |\Psi_2\rangle$.

Here we will consider the simple example presented in section 2 with the qubit starting in either $|\Psi_1\rangle=|0\rangle$ or $|\Psi_2\rangle=|+\rangle$. The way to realise in histories language the PBR argument is not unique. We will first see the simpler version, where we have a single moment of time and it is essentially the PBR argument casted into co-events language. Then we will consider an apparatus with two moments of time, that gives a better picture of the histories formulations, having given preferred status in the $\{|0\rangle,|1\rangle\}$ basis in analogy with the preferred fine grained set of histories being the configuration space basis. The latter version, involves more of the features of the co-event formulation, since both non-trivial quantum measure zero sets exist and non-classical co-events.

\subsection{Version 1}

We consider two copies (uncorrelated and with no interaction) of a qubit, that can be in either the state $|0\rangle$ or in the state $|+\rangle$. In other words the initial state is one of the following:

\bea\label{initial state}
|\Psi_1,\Psi_1\rangle=|00\rangle & , & |\Psi_1,\Psi_2\rangle=|0+\rangle\nonumber\\
|\Psi_2,\Psi_1\rangle=|+0\rangle & , & |\Psi_2,\Psi_2\rangle=|++\rangle
\eea
After preparing the initial state in one of those four states, we measure it in the $ \{|\xi_i\rangle\}$  basis given in Eq.(\ref{basis}). We label history $ h_i$ the one that the system is found in state $|\xi_i\rangle$. In other words, for example, $h_1$ is the history that the system starts from one the four states of Eq. (\ref{initial state}) and then is found being in the $|\xi_1\rangle$ state. The quantum measure, and thus the possible co-events are different for each possible initial state.

For initial state $|\Psi_1,\Psi_1\rangle=|00\rangle$, the quantum measure of different fine grained histories is the following:

\beq\mu_{11}(h_1)=0,\mu_{11}(h_2)=1/4,\mu_{11}(h_3)=1/2,\mu_{11}(h_4)=1/4\eeq
Where the subscript at the quantum measure, denotes that it corresponds to the initial state $|\Psi_1,\Psi_1\rangle$. Note that this is a trivial histories space, since it has only a single moment of time. Alternative histories, decohere, since they lie at the final moment of time, and thus the quantum measure is additive and fully given by the quantum measure on the fine grained histories. If there were more moments of time, in order to fully characterise the quantum measure, one could give the amplitudes for different histories as we will do at version 2.

There is only one quantum measure zero set, which is the single history $\{h_1\}$ and thus the potential co-events are all classical. The set of possible co-events for the initial state $|00\rangle$ are

\beq \mathcal C_1=\{\{h_2\},\{h_3\},\{h_4\}\} \eeq
For initial state $|\Psi_1,\Psi_2\rangle=|0+\rangle$, the quantum measure of different fine grained histories is the following:

\beq\mu_{12}(h_1)=1/4,\mu_{12}(h_2)=0,\mu_{12}(h_3)=1/2,\mu_{12}(h_4)=1/4\eeq
And the set of possible co-events for the initial state $|0+\rangle$ are

\beq \mathcal C_2=\{\{h_1\},\{h_3\},\{h_4\}\} \eeq
For initial state $|\Psi_2,\Psi_1\rangle=|+0\rangle$, the quantum measure of different fine grained histories is the following:

\beq\mu_{21}(h_1)=1/4,\mu_{21}(h_2)=1/2,\mu_{21}(h_3)=0,\mu_{21}(h_4)=1/4\eeq
And the set of possible co-events for the initial state $|+0\rangle$ are

\beq \mathcal C_3=\{\{h_1\},\{h_2\},\{h_4\}\} \eeq
For initial state $|\Psi_2,\Psi_2\rangle=|++\rangle$, the quantum measure of different fine grained histories is the following:

\beq\mu_{22}(h_1)=1/2,\mu_{22}(h_2)=1/4,\mu_{22}(h_3)=1/4,\mu_{22}(h_4)=0\eeq
And the set of possible co-events for the initial state $|++\rangle$ are

\beq \mathcal C_4=\{\{h_1\},\{h_2\},\{h_3\}\} \eeq
The important thing to notice here, is that there is no common co-event for these four different initial
states, namely:

\beq\mathcal C_1 \bigcap \mathcal C_2\bigcap \mathcal C_3\bigcap \mathcal C_4=\emptyset\eeq
This implies that there is no potential reality, that is compatible with all four above initial states. Carrying out the measurement suggested in PBR argument, leads us to exclude one of the four initial states as the initial state of the system. Which one is excluded, depends on the outcome of the measurement. This is close, but not quite the same, as saying that we can definitely distinguish between the state $|\Psi_1\rangle$ and $|\Psi_2\rangle$. We will come back to this point in section 5 in the discussion.

\subsection{Version 2}

Since the quantum measure is affected from the particulars of the measurements/histories considered, one could expect that realising the above set up in a different way could affect the conclusion in relation with the possible co-events. Since in histories formulation, preferred status is given to the ``actual'' fine grained histories that correspond to paths at the (extended) configuration space, one could see that the analogous thing to do for a qubit, is to measure it in the $\{|0\rangle,|1\rangle\}$ basis solely. So while we start with the same four possible initial states given by Eq. (\ref{initial state}) as in version 1 we consider different histories which in other words corresponds to a more fine grained description.

We have fine grained histories, starting at one of the four initial states, then both particles are ``measured'' in the $\{|0\rangle, |1\rangle\}$ basis, and finally they are measured in the $\{|\xi_i\rangle\}$ basis.

The possible histories are labeled as $h_{00\xi_i}$ if the system (no matter which initial state we consider) is found in $|00\rangle$ and then in $|\xi_i\rangle$, $h_{10\xi_i}$ if it is in $|10\rangle$ and then in $|\xi_i\rangle$ etc. In order to find the co-events, one needs for a given initial state to compute the quantum measure and find the quantum measure zero sets. Since histories in this version do not trivially decohere (unless they end at different final time) the most convenient way to write down the quantum measure, is in terms of the amplitudes of different fine grained histories. If a set of histories ends at the same outcome at the final time, then the quantum measure is simply the mod square of the sum of the amplitudes of fine grained histories. For the fine grained histories the quantum measure is given by $\mu(h_i)=|\a(h_i)|^2$.

For initial state $|\Psi_1,\Psi_1\rangle=|00\rangle$ we have these quantum amplitudes:

\bea\a_{11}(00\xi_1)=0 & , & \a_{11}(00\xi_2)=1/2 \nonumber\\
\a_{11}(00\xi_3)=1/2 & , & \a_{11}(00\xi_4)=1/\sqrt2\eea
and zero amplitude for all the other histories. Here, too, there are no non-trivial quantum measure zero sets. We therefore have only three classical co-events

\beq\mathcal C_1=\{\{h_{00\xi_2}\},\{h_{00\xi_3}\},\{h_{00\xi_4}\}\}\eeq
Important to note here that there is no co-event ending at $\xi_1$.

For initial state $|\Psi_1,\Psi_2\rangle=|0+\rangle$ we have these quantum amplitudes:

\bea\a_{12}(00\xi_1)=0 & , & \a_{12}(00\xi_2)=\frac1{2\sqrt2} \nonumber\\
\a_{12}(00\xi_3)=\frac1{2\sqrt2} & , & \a_{12}(00\xi_4)=1/2\nonumber\\
\a_{12}(01\xi_1)=1/2 & , & \a_{12}(01\xi_2)=-\frac1{2\sqrt2}\nonumber\\
\a_{12}(01\xi_3)=\frac1{2\sqrt2} & , & \a_{12}(01\xi_4)=0\nonumber\\
\a_{12}(10\xi_i)=0 & , & \a_{12}(11\xi_i)=0\eea
In this case, we see that there is one non-trivial quantum measure zero set. The set $\{h_{00\xi_2},h_{01\xi_2}\}$ has quantum measure zero, while the fine grained histories have both quantum measure $1/8$. Here again we have the following allowed co-events:

\beq\mathcal C_2=\{\{h_{00\xi_3}\},\{h_{00\xi_4}\},\{h_{01\xi_1}\}, \{h_{01\xi_4}\}\}\eeq
And we should note, that none of these co-events, ends at $\xi_2$.

For initial state $|\Psi_2,\Psi_1\rangle=|+0\rangle$ we have these quantum amplitudes:

\bea\a_{21}(00\xi_1)=0 & , & \a_{21}(00\xi_2)=\frac1{2\sqrt2} \nonumber\\
\a_{21}(00\xi_3)=\frac1{2\sqrt2} & , & \a_{21}(00\xi_4)=1/2\nonumber\\
\a_{21}(10\xi_1)=1/2 & , & \a_{21}(10\xi_2)=\frac1{2\sqrt2}\nonumber\\
\a_{21}(10\xi_3)=-\frac1{2\sqrt2} & , & \a_{21}(10\xi_4)=0\nonumber\\
\a_{21}(01\xi_i)=0 & , & \a_{21}(11\xi_i)=0\eea
we see that here too there is one non-trivial quantum measure zero set. The set $\{h_{00\xi_3},h_{10\xi_3}\}$ has quantum measure zero, while the fine grained histories have both quantum measure $1/8$. We have the following allowed co-events:

\beq\mathcal C_3=\{\{h_{00\xi_2}\},\{h_{00\xi_4}\},\{h_{10\xi_1}\}, \{h_{10\xi_2}\}\}\eeq
And we should note, that none of these co-events, ends at $\xi_3$.

Finally, for initial state $|\Psi_2,\Psi_2\rangle=|++\rangle$ we have these quantum amplitudes:

\bea\a_{22}(00\xi_1)=0 & , & \a_{22}(00\xi_2)=1/4 \nonumber\\
\a_{22}(00\xi_3)=1/4 & , & \a_{22}(00\xi_4)=\frac1{2\sqrt2}\nonumber\\
\a_{22}(01\xi_1)=\frac1{2\sqrt2} & , & \a_{22}(01\xi_2)=-1/4\nonumber\\
\a_{22}(01\xi_3)=1/4 & , & \a_{22}(01\xi_4)=0\nonumber\\
\a_{22}(10\xi_1)=\frac1{2\sqrt2} & , & \a_{22}(10\xi_2)=1/4\nonumber\\
\a_{22}(10\xi_3)=-1/4 & , & \a_{22}(10\xi_4)=0\nonumber\\
\a_{22}(11\xi_1)=0 & , & \a_{22}(11\xi_2)=1/4\nonumber\\
\a_{22}(11\xi_3)=1/4 & , & \a_{22}(11\xi_4)=-\frac1{2\sqrt2}\eea
We have several non-trivial quantum measure zero sets, namely: $\{h_{00\xi_4},h_{11\xi_4}\}$, $\{h_{00\xi_2},h_{01\xi_2}\}$ , $\{h_{01\xi_2},h_{10\xi_2}\}$ , $\{h_{01\xi_2},h_{11\xi_2}\}$ , $\{h_{00\xi_3},h_{10\xi_3}\}$ , $\{h_{01\xi_3},h_{10\xi_3}\}$ and $\{h_{10\xi_3},h_{11\xi_3}\}$.

The allowed co-events are more complicated in this case and there exist some non-classical co-events (corresponding to pairs of histories). There are two classical co-events ending at $\xi_1$, three co-events consisting of pairs of histories, ending at $\xi_2$ and the same at $\xi_3$ while there is no co-event ending at $\xi_4$. In particular, the co-events are:

 \bea\mathcal C_4 &= & \{\{h_{01\xi_1}\},\{h_{10\xi_1}\},\{h_{00\xi_2},h_{10\xi_2}\}, \{h_{00\xi_2},h_{11\xi_2}\},\{h_{10\xi_2},h_{11\xi_2}\},\nonumber\\ & &\{h_{00\xi_3},h_{01\xi_3}\}, \{h_{00\xi_3},h_{11\xi_3}\},\{h_{01\xi_3},h_{11\xi_3}\}\}\eea

The important thing to conclude from this analysis is that there is no common co-event for these different initial
states either:

\beq\mathcal C_1 \bigcap \mathcal C_2\bigcap \mathcal C_3\bigcap \mathcal C_4=\emptyset\eeq
Thus we can see that if the ontology of quantum theory is that of a co-event, we deduce that there is no conceivable reality that is compatible with the state being all four states $|00\rangle, |0+\rangle, |+0\rangle, |++\rangle$. This result, that was also present in version 1, is therefore maintained even if we further fine-grain the system. We should note that in this version of the example, we had more than one moments of times and there were non-trivial quantum measure zero sets that also lead to having some possible non-classical co-events. All this complication, did not affect the above conclusion. However, to make the final step, and conclude that we can always distinguish between distinct state-vectors, as the PBR arguments claims, we need to review and examine the assumptions made\footnote{Here we have presented the analysis for the specific example of the qubit in those two states. As in the original paper \cite{PBR} it can be extended for arbitrary states.}.

\section{Assumptions and discussion}

In the end of Section \ref{section PBR}, the assumptions of this theorem were briefly stated. The first assumption was that one can prepare a state of a system in isolation from the rest of the universe and its individual well defined physical properties depend only on this state and not in any sense from the rest of the universe, and the second that constructing multiple unrelated copies of the system is possible. The third was that the outcomes of the measuring deices respond solely to physical properties of the measured systems.

The first two assumptions are not (at least obviously) valid in general if one uses the Feynman path integral and the quantum measure. For co-events, for example, we know it is not true. The property that forbids one to take these two assumptions, is how the quantum measure of a composite system relates to the quantum measure of individual subsystems.

The quantum measure of a composite system is not in general the product of the quantum measures of the individual subsystems, even if there is no interaction between the subsystems. It is not even clear, how one would define the composite quantum measure if he was simply given the quantum measure of the subsystems.

One of the reasons for this failure, is related to the fact that to each quantum measure there correspond infinite different decoherence functionals. In particular, any two decoherence functionals that differ only in their complex part, give rise to the same quantum measure. While this does not affect the predictions for a single system, no matter the approach one takes (decoherent histories or co-events), it does affect considerations of composite systems.
If one is given the decoherence functional of a single system, there is a preferred way to construct the composite (two copies of) system decoherence functional, by simply taking the product of the two individual decoherence functionals. However, this also leads to some paradoxes.

In particular, we could have two uncorrelated and non-interacting systems, that all the individual histories are possible, i.e. no history has zero amplitude and moreover no coarse-grained history of the individual subsystems has quantum measure zero. By considering the quantum measure that arises from taking the product decoherence functional, we now have some coarse grained histories having quantum measure zero, and therefore not all combinations of individual outcomes are allowed. Moreover, this property arises from simply considering the two uncorrelated, non-interacting systems together.

The reason that this problem that appears for sets with quantum measure zero is important, is twofold. First, because sets with quantum measure zero, decohere with their negation, as we show earlier, and thus necessarily belong to one decoherent set. The second, is because those sets are used in the co-event formulation in order to find the potential co-event.

A simple example of what could go wrong is given here. Assume we have a decoherence functional given by:

\beq D_A=\frac12\left(\begin{array}{c c}1&i\\ -i&1\end{array}\right)\eeq
In this example, no history is precluded, since there exist no quantum measure zero set. History $h_1$ corresponds to the $(1,1)$ entry of the matrix and $h_2$ to the $(2,2)$. One can also note, that for this system there is no non-trivial decoherent set since we adopt the medium decoherence condition that requires both the real and imaginary part of the off-diagonal parts of the decoherence functional to vanish\footnote{This particular example, is weakly decoherent, but one can construct more complicated examples that even this is not true.}. Consider two identical systems and take their deoherence functional be their product:

\beq D_{AB}=\frac14\left(\begin{array}{c c c c}1&i&i&-1\\ -i&1&1&i\\-i&1&1&i\\-1&-i&-i&1\end{array}\right)\eeq
We can now see that there is a nontrivial quantum measure zero set.

\beq\mu(\{h_{11},h_{22}\})=0\nonumber\eeq
We can conclude from that that the coarse-grained history $\{h_{11},h_{22}\}$ is not possible. A subset of this history, and thus also precluded, is the history where both subsystems are at $h_1$. Similarly the history that both subsystems are at $h_2$ is subset of $\{h_{11},h_{22}\}$ and thus precluded. However, no such preclusion was possible if we were simply looking the single systems decoherence functionals. One can claim that there is an (anti)correlation of the two systems, even though there is no entanglement and interaction.

Another aspect of this property can be seen, if we note that for the composite system, there is a non-trivial decoherent set, that follows from the existence of the zero set. The partition of $\Omega=\{A,B\}$ where $A=\{h_{11},h_{22}\}$ and $B=\{h_{12},h_{21}\}$ is a decoherent set. For the individual systems however, there was no such set. Note that, while it is true that the product of two non-interacting (medium) decoherent sets of histories give rise to a decoherent set of histories, the converse is not true. We can have a decoherent set of histories on the product system of two non-interacting subsystems, with no analogue for the individual subsystems. This observation, highlights the failure of the assumptions of the PBR argument for histories.

A side-note, here is that it was already known that there are issues with composite systems if one considers the decoherence functional. For example in the consistent histories framework, Diosi in \cite{Diosi} showed that if one requires weak decoherence, i.e. that only the real part of the decoherence functional needs to vanish, then one is lead to contradictions by considering non-interacting composite systems. In particular, the weak decoherence condition might hold for subsystems but not for the total composite system\footnote{In his work on extending the probabilities \cite{Ha2008}, Hartle has also point out this strange property for his extended probabilities. According to his work, however, when restricted to ``settleable'' questions the probabilities of composite systems, behave classically.}. This was probably the strongest reason of adopting the medium (or stronger) decoherence condition for the decoherent histories approach.

From the latter observation, one could be tempted to conclude, that issues with composite systems, arise from the complex part of the decoherence functional and that a possible restriction to real decoherence functional could resolve them\footnote{Of course one would need to show that the real decoherence functionals give rise to all the properties one would expect from experiments.}. However, this is also \emph{not} true. One can construct a purely real decoherence functional and when considering a composite system, still getting new quantum measure zero sets and their interpretational consequences\footnote{However, the examples with purely real decoherence functional, involve higher cardinality histories space $\Omega$, since no such example exist in the 2 histories space mentioned above.}. The analysis of this property is the subject of an ongoing work \cite{DoSoWa}.

To this end, we must stress that there is a special type of correlation between systems in histories formulations that is not related to entanglement. This precise property, forbids one to make the final step at the PBR argument. To recall the full argument, it was first shown that if one measures in the $\{|\xi_i\rangle\}$ basis, any possible outcome is consistent with the system having started with only three of the total four possible initial states of the composite system $|\Psi_1,\Psi_1\rangle$, $|\Psi_1,\Psi_2\rangle$, $|\Psi_2,\Psi_1\rangle$ and $|\Psi_2,\Psi_2\rangle$. e.g. if $\xi_1$ is the outcome of the measurement, we know definitely the system did not start at $|\Psi_1,\Psi_1\rangle$, but could have started form any of the other three composite states. The second part of the argument, was to relate this with properties of the single system. In particular, they concluded that there is no outcome that can be consistent with both systems being both $|\Psi_1\rangle$ and $|\Psi_2\rangle$. If $|\Psi_1\rangle$ and $|\Psi_2\rangle$ were merely overlapping distributions of some underlying potential realities, then with some non-zero probability the reality that is actually realised, would lie in the overlapping part and thus, in these cases, any possible measurement would not be able to tell which was the initial state. The initial state would no longer have any ontological status.

We need to stress here, that the second part of the argument, lies on the assumption, that the joint distribution of the two single systems is merely the product of the distributions of the individual systems (for unrelated, disentangled, non-interacting systems). However, as we analysed above, this is not true for the histories and the quantum measure of composite systems, and it is not at all clear if and how can this argument be completed.

The conclusion of the PBR argument, namely that state-vectors cannot be interpreted statistically, could still hold for co-events (in the sense that the set of allowed co-events for different initial states are completely disjoint), but it is not proven with this thought experiment. For example, it would be interesting to explore other possibilities such as extending alternative proposals to the PBR argument (one very recent one is  \cite{Patra}) for the co-events formulation.

In our case, for histories formulations and the co-events in particular, we actually expect it to be true. By extending the histories for sufficient time and sufficiently fine-grained, we conjecture (and give further evidence for it) that the set of possible co-events for any separate state-vectors would be distinct. This would suggest, that if we observe with sufficient detail the system for long enough, we will be able to deduce uniquely which was the initial state, provided that it was a pure state. Evidence that this conjecture holds, is given in the appendix, where we show that indeed for essentially \emph{any} two distinct initial state-vectors $|\Phi_1\rangle$ and $|\Phi_2\rangle$ of a qubit (2-dim Hilbert space), we can explicitly construct sufficiently fine grained histories, such that the set of co-events $\mathcal C_1$ and $\mathcal C_2$ have no common elements ($\mathcal C_1 \cap\mathcal C_2=\emptyset$). Distinguishing between two states is something that always concerns the 2-dim subspace of the Hilbert space that is spanned by those states (as was stressed in \cite{PBR} and explained in detail in the appendix). Therefore the evidence given in the appendix, that the conjecture holds, is very strong, failing to form a full proof due to some special cases and some extra attention needed for the infinite dimensional case.

\section{Summary and conclusion}

We first reviewed the PBR argument for why a statistical interpretation of the quantum state is not possible. We introduced the histories formulations and in particular the decoherent histories approach and the co-event formulation. The reason to examine the PBR argument for histories is twofold. First, in order to make contact with standard one-moment-of-time quantum theory and compare the role of the state. Second reason, is related with retrodiction. While it is not clear that one can determine uniquely to arbitrary precision which one of the potential realities is eventually realised,  it is clear that if two different initial state-vectors can give rise even to one common reality, there is no hope of one being able to distinguish between the two situations with certainty. Even speaking about which was the initial state-vector, in this case, is meaningless. Note that histories formulations, are frequently applied to the field of quantum cosmology where the initial state and the ability to retrodict are very important.

We then showed that the PBR argument applies to histories formulations looking in section 4, at two versions of the specific example considered in section 2. One is always able to distinguish between the four (at the example) state-vectors $|\Psi_1,\Psi_1\rangle$, $|\Psi_1,\Psi_2\rangle$ , $|\Psi_2,\Psi_1\rangle$ and $|\Psi_2,\Psi_2\rangle$. However one cannot make the further step and conclude that we can distinguish between single system state-vectors $|\Psi_1\rangle$ and $|\Psi_2\rangle$, as we saw in section 5. The reason is that some of the assumptions of PBR, do not hold in histories formulations. In particular, there is a strange correlation, not related with entanglement, between sub-systems that the decoherence functional of composite systems has. This property needs to be further examined and understood \cite{DoSoWa}. The author however conjectures, that it would still be possible to distinguish between different initial states, if one extends the histories suitably, for the co-event formulation. Evidence for the validity of the conjecture is given in the appendix, where it is proven explicitly that this is the case for 2-dim Hilbert space except some very special case, and that it can be generalised to higher (finite) dimensions. To sum up, the conclusion of the PBR argument is expected to be valid for histories as well, but it cannot be proven with their gedanken experiment.

\paragraph{Acknowledgments:} The author is very grateful to Rafael Sorkin, for bringing the problem to his attention, many discussions and reading and commenting an earlier draft. He acknowledges the COST Action MP1006 ``Fundamental Problems in Quantum Physics'' and also the Perimeter Institute for Theoretical Physics, Waterloo, Canada, for hospitality while carrying out part of this work.

%The author wants to thank Rafael Sorkin for bringing the problem to his attention and for many discussions in relation with the understanding the conclusion. ... COST MP1006... Perimeter Institute?

\section*{Appendix}

In this appendix, we will attempt to prove the conjecture made in the text, that the allowed co-events for two different state-vectors, are disjoint. We will use the example of a qubit. However, the importance of this example, is much greater. When comparing two state-vectors\footnote{For simplicity, of a finite dimensional Hilbert space. For infinite dimensions more care is needed.}, we can do so, by restricting attention to the 2-dimensional subspace that is spanned by the two state-vectors. While it is true, that many results hold in 2-dimensions and not for higher dimensions, the comparison of two pure states, is not one of them. The reason this is the case is because if one has two state-vectors of higher dimensions $|\Phi_1\rangle$, $|\Phi_2\rangle$ $\in \mathcal{H}$, he can find the two-dimensional subspace that is spanned by those two vectors $\mathcal{H}_{\Phi_1,\Phi_2}=\textrm{span}(\{|\Phi_1\rangle,|\Phi_2\rangle\})$. One can then choose $|\Phi_1\rangle=|0\rangle$ and define $|1\rangle\in\mathcal{H}_{\Phi_1,\Phi_2}$ such that it is orthogonal to $|0\rangle$, i.e. $\langle 0|1\rangle=0$. Then, we can express $|\Phi_2\rangle= \cos \theta|0\rangle + \sin \theta |1\rangle$ for some angle $\theta$, and without loss of generality, we can proceed as if our initial states $|\Phi_1\rangle,|\Phi_2\rangle$ were 2-dimensional.

Our attempt to prove the conjecture, consist of the explicit construction of the sets of co-events corresponding to two different initial states, that are non-orthogonal. By choosing finer or coarser grained histories (i.e. by having more moments-of-time) we can construct finer or coarser sets of co-events. The direction of proof we will follow, is to exploit some property that certain coarse-graining of histories have, namely the existence of zero covers \cite{coevents_review}. This simplifies considerably the analysis and allows us to find a suitable coarse-graining that distinguishes the states with only 3-moments-of-time, and therefore $2^3$ different histories for almost all cases. Unfortunately, as we will see below, for the very special case that the angle between the two states we want to distinguish is $\tan\theta=\pm1/3$, this strategy is not successful. To fully prove the conjecture, one needs to consider different coarse-grainings with more moments-of-time (and therefore exponentially more possible histories). With the current development of the co-event formulation, its extremely difficult to compute the possible co-events, as the moments of time increase. Moreover, the technical trick used to prove the conjecture for all the other cases (the use of that particular zero cover), cannot be used here. However, it seems very implausible, that one can distinguish between any two state-vectors in this formalism, unless their angle is $\tan^{-1}(\pm1/3)$. If that was the case, it would certainly be a strange property that would require further study. It is most likely that some other finer-grained description would complete the proof, but until the relevant technical methods to efficiently compute co-events for many moments-of-time appears, our claim will remain a conjecture. We now return to prove the general case.

The histories we will consider, is essentially 3-moments-of-time. We start with some initial state $|\Phi\rangle$ and then measure it three times in the basis we will give below. We assume trivial evolution (the identity), but we could easily have any Hamiltonian, and then have to choose the basis measured suitably. Given a particular Hamiltonian (non-trivial this time), one can also reproduce the result we will give, considering measurement done only in the $\{|0\rangle,|1\rangle\}$ basis, by suitably choosing the time $t_1,t_2,t_3$ that the measurements take place as we will see in the end of the appendix.

The initial state will be either $|\Phi_1\rangle= |0\rangle$ or any other state $|\Phi_2\rangle = \cos\theta|0\rangle+\sin\theta|1\rangle$. We consider the following two orthogonal bases\footnote{Note that $|\Phi_2\rangle=|\Psi_0\rangle$}:

\bea
|\Psi_0\rangle &=& \cos\theta|0\rangle+\sin\theta|1\rangle \nonumber\\
|\Psi_1\rangle &=& -\sin\theta|0\rangle+\cos\theta|1\rangle
\eea
and
\bea
|\Psi_+\rangle &=& \cos(\theta+\pi/4)|0\rangle+\sin(\theta+\pi/4)|1\rangle \nonumber\\
|\Psi_-\rangle &=& \cos(\theta-\pi/4)|0\rangle+\sin(\theta-\pi/4)|1\rangle
\eea
The histories considered will be: They start with the initial state $|\Phi_i\rangle$, and then are measured in the $\{|\Psi_+\rangle,|\Psi_-\rangle\}$ basis then in the $\{|\Psi_0\rangle,|\Psi_1\rangle\}$ basis and then again in the $\{|\Psi_+\rangle,|\Psi_-\rangle\}$. We will label the histories depending on the outcome of each measurement in the following way (measurements are from right to left):

\bea h_1 = (\Psi_+\Psi_0\Psi_+),\quad h_2=(\Psi_+\Psi_1\Psi_+),\quad h_3=(\Psi_+\Psi_0\Psi_-), \quad h_4=(\Psi_+\Psi_1\Psi_-)\nonumber\\
h_5 = (\Psi_-\Psi_0\Psi_+),\quad h_6=(\Psi_-\Psi_1\Psi_+),\quad h_7=(\Psi_-\Psi_0\Psi_-), \quad h_8=(\Psi_-\Psi_1\Psi_-)
\eea
Histories $h_1,h_2,h_3$ and $h_4$ end at final time in the $|\Psi_+\rangle$ while $h_5,h_6,h_7$ and $h_8$ end in $|\Psi_-\rangle$. We compute the amplitudes of histories for $|\Phi_1\rangle=|0\rangle$ (the subscript at the amplitudes $\a_1$ signifies that it correspond to initial state $|\Phi_1\rangle$):

\bea\label{appendix_amplitudes_1}
\a_1(h_1)= 1/2\cos(\theta+\pi/4)&,& \a_1(h_2)=1/2\cos(\theta+\pi/4)\nonumber\\
\a_1(h_3)= 1/2\cos(\theta-\pi/4)&,& \a_1(h_4)=-1/2\cos(\theta-\pi/4)\nonumber\\
\a_1(h_5)= 1/2\cos(\theta+\pi/4)&,& \a_1(h_6)=-1/2\cos(\theta+\pi/4)\nonumber\\
\a_1(h_7)= 1/2\cos(\theta-\pi/4)&,& \a_1(h_8)=1/2\cos(\theta-\pi/4)
\eea
The only zero quantum measure sets are the $\{h_3,h_4\}$ and $\{h_5,h_6\}$ for a general angle $\theta$.

Here we should note that there are other zero quantum measure sets only in the cases where $\theta=0$, that reduces to $|0\rangle$ which we will see below, and for $\tan\theta=\pm 1/3$ which is the exceptional case mentioned earlier that prevents us from providing a full proof of the conjecture. For these very special cases, the fine grained description we used here is not sufficient to prove the conjecture, and further fine graining (measurements) are required.

Returning to the general $\theta$ case, we have 4 classical co-events and the set of allowed co-events are:

\beq\mathcal{C}_1=\{\{h_1\},\{h_2\},\{h_7\},\{h_8\}\}\eeq
For $|\Phi_2\rangle = \cos\theta|0\rangle+\sin\theta|1\rangle$ the amplitudes are (note that they are independent of $\theta$):

\bea\label{appendix_amplitudes_2}
\a_2(h_1)=\frac1{2\sqrt2} &,& \a_2(h_2)=\frac1{2\sqrt2}\nonumber\\
\a_2(h_3)=\frac1{2\sqrt2} &,& \a_2(h_4)=-\frac1{2\sqrt2}\nonumber\\
\a_2(h_5)=\frac1{2\sqrt2} &,& \a_2(h_6)=-\frac1{2\sqrt2}\nonumber\\
\a_2(h_7)=\frac1{2\sqrt2} &,& \a_2(h_8)=\frac1{2\sqrt2}
\eea
The subscript $\a_2$ signifies that that the initial state is $|\Phi_2\rangle$. The sets with quantum measure zero are: \beq\{h_1,h_4\},\{h_2,h_4\},\{h_3,h_4\},\{h_5,h_6\},\{h_6,h_7\},\{h_6,h_8\}\eeq
We see that all fine grained histories are contained in one quantum measure zero set and thus there are no classical co-events. Only pairs of histories are allowed, and we have 6 potential co-events:

\beq\mathcal{C}_2=\{\{h_1,h_2\},\{h_1,h_3\},\{h_2,h_3\},\{h_5,h_7\},\{h_5,h_8\},\{h_7,h_8\}\}\eeq
It is easy to see that $\mathcal{C}_1\cap\mathcal{C}_2=\emptyset$. This is general for an arbitrary $\theta$ (other than a very special case mentioned above), and thus any two state-vectors of a qubit give rise to completely disjoint set of potential co-events and thus correspond to different ontology, in the sense discussed in the main text.

We now return at the earlier remark, that all of the above can be re-expressed in terms of measurements in the $\{|0\rangle,|1\rangle\}$ basis given a Hamiltonian, if we suitably choose the times that the measurements take place. If for example we this Hamiltonian

\beq H=\left(\begin{array}{c c}1&i\\ -i&1\end{array}\right)\eeq
It gives rise to the following unitary evolution

\beq U(t)=\exp(-it)\left(\begin{array}{c c}\cos t&\sin t\\ -\sin t&\cos t\end{array}\right)\eeq
We choose to measure at $t_1=(\theta-\pi/4)$ and at $t_2=\theta$ and finally at $t_3=(\theta+7\pi/4)$, always in the $\{|0\rangle,|1\rangle\}$ basis. It is easy to calculate that the amplitudes we get for these histories for any of the two initial states, are exactly the same as the ones we calculated earlier in the appendix in Eqs. (\ref{appendix_amplitudes_1} , \ref{appendix_amplitudes_2}), with the following adjustments. (a) In the labeling we replace $\Psi_+$ and $\Psi_1$ with $1$ while we replace $\Psi_-$ and $\Psi_0$ with $0$ (i.e. the history $h_3$ for example, that was $(\Psi_+\Psi_0\Psi_-)$ is now $(100)$) and (b) there is an overall factor of $\exp (-i(\theta-\pi/4))$ in the amplitudes of \emph{all} histories, which however, does not affect the quantum measure. Since the quantum measure is the same, it follows that set of allowed co-events is also the same.

\end{document}